# A Fuzzy Based Conceptual Framework For Career Counselling

Raj Kishor Bisht

Department of Applied Sciences
Amrapali Institute of Technology and Sciences,
Haldwani, Uttarakhand, India
bishtrk@gmail.com

## ABSTRACT

*Career guidance for students, particularly in rural areas is a challenging issue in India. In the present era of digitalization, there is a need of an automated system that can analyze a student for his/her capabilities, suggest a career and provide related information. Keeping in mind the requirement, the present paper is an effort in this direction. In this paper, a fuzzy based conceptual framework has been suggested. It has two parts; in the first part a students will be analyzed for his/her capabilities and in the second part the available courses, job aspects related to their capabilities will be suggested. To analyze a student, marks in various subject in 10+2 standards and vocational interest in different fields have been considered and fuzzy sets have been formed. On example basis, fuzzy inference rules have been framed for analyzing the abilities in engineering, medical and hospitality fields only. In second part, concept of composition of relations has been used to suggest the related courses and jobs.*

## 1. INTRODUCTION

Since our childhood, we observe too many things from our society. We keep on searching the best suitable career for us. Certain jobs attract us due to our interest or due to their socio-economic impact in the society. In some of the cases, students get the right career after 10+2 standard but in general, a number of students are highly confused about their career. They do not have the capability to choose the right career for them; parents or the surrounding environment are also not helpful to them. Many times a student chooses a wrong career just because of the pressure of family or due to misguidance. The result of the same is the failure in that career and ultimately wastage of time and money. To the solution of this problem, we need an automated system that can analyze a student and suggest him/her a career according to his/her capabilities particularly for those students who live in remote areas. The automated system should not only assist a student for suitable career but it should also provide a complete guideline about the related courses, available institutes within certain area and available jobs.

Though there are counsellors working in this area but these are available mainly in cities. For the students living in remote areas, an automated system is quite useful. So far some of the efforts that have been made in this direction are as follows: Bresfelean and Ghisoiu [1] discussed various issues related to decision support system for higher education. Kostoglou et al [3] reviewed existing web-based decision support systems and elaborated some main element in the analysis and the design of these systems for higher education in Greece. Mundra et al [4] proposed an education decision support system model that includes the components; user interface, inference engine and knowledge base. Oladokun and Oyewole [5] proposed a fuzzy inference based decision support system for students facing problems in choosing courses in university by taking example of Nigerian university system.

In section 2, mathematical formulations of different phases have been defined by taking example of certain career fields. Some experimental results have been demonstrated on example basis. Finally section 3 summarizes the present work.

## 2. MATHEMATICAL FORMULATION

The proposed system consists two phases described as follows:

### 2.1 Phase 1

In this phase a student will be analyzed for his/her subjective capabilities and vocational interests. After analyzing these capabilities, a student will be provided a direction to choose the field of his potential. To analyze subjective capabilities, subjects at 10+2 levels can be considered like Physics, Chemistry, Mathematics, Biology, Home Science, Commerce, Social Science, Arts etc. For vocational interests, we can utilize some available standards test batteries or we can create them. The vocational test battery [2] checks the vocational interest in scientific, executive, agricultural, commercial, House hold, social, literary, artistic etc. The standard test batteries analyze a person and provide decisions in various categories like poor, below average, average, good, excellent etc. Here we are concerned with last three categories only.

For subjective interest, we assume that the maximum marks in each subject are 100. Let $x$ be the actual marks in a subject. To decide the subjective capabilities of a student, the three different categories are defined as follows:

Average: $$\mu_a(x) = \begin{cases} \dfrac{x-40}{15} & for \quad 40 \leq x \leq 55 \\ 1 & for \quad 55 < x \leq 60 \\ 0 & otherwise \end{cases}, \qquad (1)$$

Good: $$\mu_g(x) = \begin{cases} \dfrac{x-55}{15} & for \quad 55 \leq x \leq 70 \\ 1 & for \quad 70 < x \leq 75 \\ 0 & otherwise \end{cases}, \qquad (2)$$

Excellent: $$\mu_e(x) = \begin{cases} \dfrac{x-70}{20} & for \quad 70 \leq x \leq 90 \\ 1 & for \quad x > 90 \\ 0 & otherwise \end{cases} \qquad (3)$$

The graphical representation of the three categories is shown in figure 1.

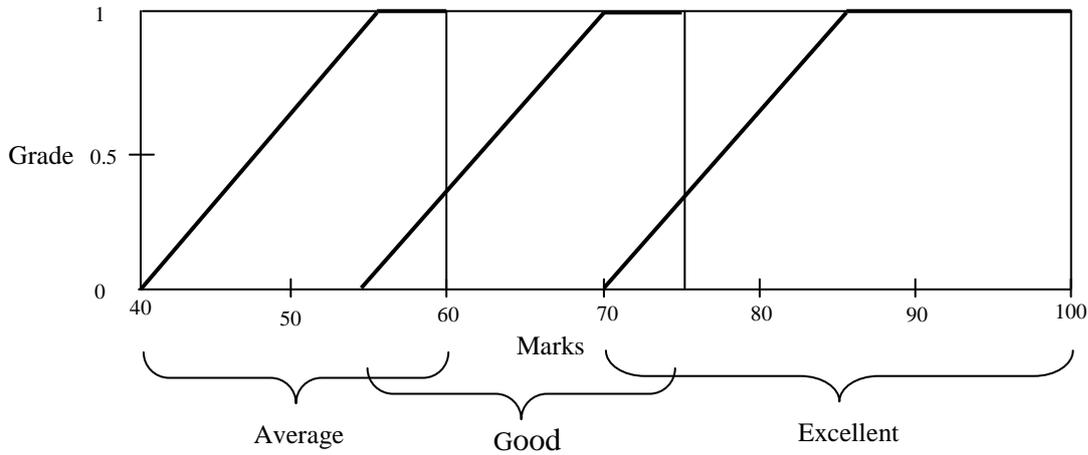

Figure 1. Membership functions for different categories

With help of equations (1),(2) and (3), we get the grade of membership of subject ability to a particular category which shows the grade of belongingness of the particular subjective ability of a student to a particular category.

In order to summarize the subjective ability and vocational interest in different career fields, a number of fields may be selected like, Engineering, Medical, Hospitality, Commerce, and Literature etc. To interpret the student's ability in a particular field, two categories have been formed namely, Excellent and Good.

Here the rules have been defined for engineering, medical and hospitality fields on example basis. For engineering some importance is given to Mathematics and Physics as a number of engineering branches are directly related to these subjects. Further weighted average has been calculated to find the grades in the two categories. Let $x_m, x_p$ and $x_c$ denote the marks in Mathematics, Physics and Chemistry respectively and $\mu_e(x), \mu_g(x)$ and $\mu_a(x)$ denote the grades of membership of $x$ in 'Excellent', 'Good' and 'Average' categories respectively. Following fuzzy rules are defined to aggregate the scores and interpret the result for engineering field:

Fuzzy inference rules for Engineering:

(a) If Mathematics is Excellent and Physics is Excellent and Chemistry is Excellent, then Engineering is Excellent. $E_1(x) = \frac{1}{3}[\mu_e(x_m) + \mu_e(x_p) + \mu_e(x_c)]$

(b) If Mathematics is Excellent and Physics is Excellent and Chemistry is Good, then Engineering is Excellent. $E_2(x) = 0.4\mu_e(x_m) + 0.4\mu_e(x_p) + 0.2\mu_g(x_c)$

(c) If Mathematics is Excellent and Physics is Good and Chemistry is Excellent, then Engineering is Excellent. $E_3(x) = 0.4\mu_e(x_m) + 0.2\mu_g(x_p) + 0.4\mu_e(x_c)$

(d) If Mathematics is Good and Physics is Excellent and Chemistry is Excellent then Engineering is Excellent. $E_4(x) = 0.2\mu_g(x_m) + 0.4\mu_e(x_p) + 0.4\mu_e(x_c)$

(e) If Mathematics is Excellent and Physics is Good and Chemistry is Good then Engineering is Good. $G_1(x) = 0.4\mu_e(x_m) + 0.3\mu_g(x_p) + 0.3\mu_g(x_c)$

(f) If Mathematics is Good and Physics is Excellent and Chemistry is Good then Engineering is Good. $G_2(x) = 0.3\mu_g(x_m) + 0.4\mu_e(x_p) + 0.3\mu_g(x_c)$

(g) If Mathematics is Good and Physics is Good and Chemistry is Excellent then Engineering is Good. $G_3(x) = 0.3\mu_g(x_m) + 0.3\mu_g(x_p) + 0.4\mu_e(x_c)$

(h) If Mathematics is Good and Physics is Good and Chemistry is Good then Engineering is Good. $G_4(x) = \frac{1}{3}[\mu_g(x_m) + \mu_g(x_p) + \mu_g(x_c)]$.

(i) If Mathematics is Excellent and Physics is Good and Chemistry is Average then Engineering is Good. $G_5(x) = 0.4\mu_e(x_m) + 0.35\mu_g(x_p) + 0.25\mu_a(x_c)$

(j) If Mathematics is Good and Physics is Excellent and Chemistry is Average then Engineering is Good. $G_6(x) = 0.35\mu_g(x_m) + 0.4\mu_e(x_p) + 0.25\mu_a(x_c)$

(k) If Mathematics is Good and Physics is Good and Chemistry is Average then Engineering is Good. $G_7(x) = 0.4\mu_g(x_m) + 0.4\mu_g(x_p) + 0.2\mu_a(x_c)$

The grade of membership for engineering field is assigned by aggregating the grades obtained by applying various rules as follows:

Excellent: $E_e(x) = Average\{E_i(x) : 1 \leq i \leq 4\}$, Good: $G_e(x) = Average\{G_i(x) : 1 \leq i \leq 7\}$

Let $x_b, x_p$ and $x_c$ denote the marks in Biology, Physics and Chemistry respectively. Following fuzzy rules are defined to aggregate the scores and interpret the result for medical field:

Fuzzy inference rules for Medical:

(a) If Biology is Excellent and Physics is Excellent and Chemistry is Excellent, then Medical is Excellent. $E_1(x) = \frac{1}{3}[\mu_e(x_b) + \mu_e(x_p) + \mu_e(x_c)]$

(b) If Biology is Excellent and Physics is Excellent and Chemistry is Good, then Medical is Excellent. $E_2(x) = 0.4\mu_e(x_b) + 0.4\mu_e(x_p) + 0.2\mu_g(x_c)$

(c) If Biology is Excellent and Physics is Good and Chemistry is Excellent, then Medical is Excellent. $E_3(x) = 0.4\mu_e(x_b) + 0.2\mu_g(x_p) + 0.4\mu_e(x_c)$

(d) If Biology is Good and Physics is Excellent and Chemistry is Excellent then Medical is Excellent. $E_4(x) = 0.2\mu_g(x_b) + 0.4\mu_e(x_p) + 0.4\mu_e(x_c)$

(e) If Biology is Excellent and Physics is Good and Chemistry is Good then Medical is Good. $G_1(x) = 0.4\mu_e(x_b) + 0.3\mu_g(x_p) + 0.3\mu_g(x_c)$

(f) If Biology is Good and Physics is Excellent and Chemistry is Good then Medical is Good. $G_2(x) = 0.3\mu_g(x_b) + 0.4\mu_e(x_p) + 0.3\mu_g(x_c)$

(g) If Biology is Good and Physics is Good and Chemistry is Excellent then Medical is Good. $G_3(x) = 0.3\mu_g(x_b) + 0.3\mu_g(x_p) + 0.4\mu_e(x_c)$

(h) If Biology is Good and Physics is Good and Chemistry is Good then Medical is Good.
$G_4(x) = \frac{1}{3}[\mu_g(x_b) + \mu_g(x_p) + \mu_g(x_c)]$.

(i) If Biology is Good and Physics is Good and Chemistry is Average then Medical is Good. $G_5(x) = 0.4\mu_g(x_b) + 0.4\mu_g(x_p) + 0.2\mu_a(x_c)$

(j) If Biology is Good and Physics is average and Chemistry is Good then Medical is Good. $G_6(x) = 0.4\mu_g(x_b) + 0.2\mu_a(x_p) + 0.4\mu_g(x_c)$

(k) If Biology is Average and Physics is Good and Chemistry is Good then Medical is Good. $G_7(x) = 0.2\mu_a(x_b) + 0.4\mu_g(x_p) + 0.4\mu_g(x_c)$

The grade of membership for medical field is assigned by aggregating the grades obtained by applying various rules as follows:

Excellent: $E_m(x) = Average\{E_i(x) : 1 \leq i \leq 4\}$, Good: $G_m(x) = Average\{G_i(x) : 1 \leq i \leq 7\}$

So far fuzzy rules have been defined for engineering and medical fields. These fields need only subjective abilities. Now another field 'Hospitality' is considered, that requires vocational interest also. For this we need to check the vocational interest of a student in house hold and the standard test batteries can be utilized for testing the ability. Let $x_{hs}$ and $x_{hh}$ denote the scores in home science and vocational interest 'house hold' respectively. Following fuzzy rules are defined to aggregate the scores and interpret the result for hospitality field:

Fuzzy inference rules for Hospitality

(a) If Home Science is Excellent and House hold is Excellent, then Hospitality is Excellent.
$E_{h1}(x) = \frac{1}{2}[\mu_e(x_{hs}) + \mu_e(x_{hh})]$

(b) If Home Science is Excellent and House hold is Good, then Hospitality is Excellent.
$E_{h2}(x) = 0.7\mu_e(x_{hs}) + 0.3\mu_g(x_{hh})$

(c) If Home Science is Good and House hold is Excellent, then Hospitality is Excellent.
$E_{h3}(x) = 0.3\mu_g(x_{hs}) + 0.7\mu_e(x_{hh})$

(d) If Home Science is Good and House hold is Good, then Hospitality is Good.
$G_{h1}(x) = 0.5\mu_g(x_{hs}) + 0.5\mu_g(x_{hh})$

(e) If Home Science is Good and House hold is Average, then Hospitality is Good.
$G_{h2}(x) = 0.7\mu_g(x_{hs}) + 0.3\mu_a(x_{hh})$

(f) If Home Science is Average and House hold is Good, then Hospitality is Good.
$G_{h3}(x) = 0.3\mu_a(x_{hs}) + 0.7\mu_g(x_{hh})$

The grade of membership for hospitality field is assigned by aggregating the grades obtained by applying various rules as follows:

Excellent: $E_h(x) = Average\{E_{hi}(x) : 1 \leq i \leq 3\}$, Good: $G_h(x) = Average\{G_{hi}(x) : 1 \leq i \leq 3\}$

### 2.2 Experimental demonstration of phase 1

For experimental purpose, we have assumed the marks of three arbitrary students. Further we have assumed the scores in vocational interest 'house hold' in the scale of 100. In table 1, we have shown the grades of different students in three categories.

Table 1. Grades of membership of students in different subjects and categories

| Student ID | Subject | Marks/ Score (x) | $\mu_a(x)$ | $\mu_g(x)$ | $\mu_e(x)$ |
|---|---|---|---|---|---|
| 1 | Maths. | 82 | 0.00 | 0.00 | 0.80 |
| 1 | Phy. | 85 | 0.00 | 0.00 | 1.00 |
| 1 | Chem. | 86 | 0.00 | 0.00 | 1.00 |
| 1 | Biology | 65 | 0.00 | 0.67 | 0.00 |
| 1 | Home Science | 56 | 1.00 | 0.07 | 0.00 |
| 1 | House hold | 60 | 1.00 | 0.33 | 0.00 |
| 2 | Maths. | 63 | 0.00 | 0.53 | 0.00 |
| 2 | Phy. | 72 | 0.00 | 1.00 | 0.13 |
| 2 | Chem. | 70 | 0.00 | 1.00 | 0.00 |
| 2 | Biology | 79 | 0.00 | 0.00 | 0.60 |
| 2 | Home Science | 60 | 1.00 | 0.33 | 0.00 |
| 2 | House hold | 58 | 1.00 | 0.20 | 0.00 |
| 2 | Maths. | 53 | 0.87 | 0.00 | 0.00 |
| 2 | Phy. | 55 | 1.00 | 0.00 | 0.00 |
| 2 | Chem. | 56 | 1.00 | 0.07 | 0.00 |
| 2 | Biology | 59 | 1.00 | 0.27 | 0.00 |
| 2 | Home Science | 72 | 0.00 | 1.00 | 0.13 |
| 2 | House hold | 76 | 0.00 | 0.00 | 0.40 |

Table 2 shows the scores of various rules and final grades of two categories for different fields based on the scores given in table 1 for different students and their grades in different subject or vocational interest.

Table 2. Grades of membership of students in different fields

| Student ID | Career Field | Categories | Rules and grades of membership | | | | | | | Final grades |
|---|---|---|---|---|---|---|---|---|---|---|
| 1 | Engg. | Excellent | A | B | C | D | | | | |
| | | | 0.93 | 0.72 | 0.72 | 0.80 | | | | 0.79 |
| | | Good | E | F | G | H | I | J | K | |
| | | | 0.32 | 0.40 | 0.40 | 0.00 | 0.32 | 0.40 | 0.00 | 0.26 |
| | Medical | Excellent | A | B | C | D | | | | |
| | | | 0.67 | 0.13 | 0.40 | 0.93 | | | | 0.53 |
| | | Good | E | F | G | H | I | J | K | |
| | | | 0 | 0.60 | 0.40 | 0.22 | 0.27 | 0.27 | 0.00 | 0.25 |
| | Hospitality | Excellent | A | B | C | | | | | |
| | | | 0 | 0.1 | 0.02 | | | | | 0.04 |
| | | Good | D | E | F | | | | | |
| | | | 0.20 | 0.35 | 0.53 | | | | | 0.36 |
| 2 | Engg. | Excellent | A | B | C | D | | | | |
| | | | 0.04 | 0.25 | 0.20 | 0.16 | | | | 0.16 |
| | | Good | E | F | G | H | I | J | K | |
| | | | 0.6 | 0.51 | 0.46 | 0.84 | 0.35 | 0.24 | 0.61 | 0.52 |
| | Medical | Excellent | A | B | C | D | | | | |
| | | | 0.24 | 0.16 | 0.44 | 0.05 | | | | 0.23 |
| | | Good | E | F | G | H | I | J | K | |
| | | | 0.84 | 0.35 | 0.46 | 0.67 | 0.40 | 0.40 | 0.80 | 0.56 |
| | Hospitality | Excellent | A | B | C | | | | | |
| | | | 0 | 0.06 | 0.1 | | | | | 0.05 |
| | | Good | D | E | F | | | | | |
| | | | 0.27 | 0.53 | 0.44 | | | | | 0.41 |
| 3 | Engg. | Excellent | A | B | C | D | | | | |
| | | | 0.00 | 0.01 | 0.00 | 0.00 | | | | 0.00 |
| | | Good | E | F | G | H | I | J | K | |
| | | | 0.02 | 0.02 | 0.00 | 0.02 | 0.25 | 0.25 | 0.30 | 0.12 |
| | Medical | Excellent | A | B | C | D | | | | |
| | | | 0.00 | 0.00 | 0.00 | 0.05 | | | | 0.01 |
| | | Good | E | F | G | H | I | J | K | |
| | | | 0.02 | 0.10 | 0.00 | 0.11 | 0.31 | 0.33 | 0.23 | 0.16 |
| | Hospitality | Excellent | A | B | C | | | | | |
| | | | 0.27 | 0.09 | 0.58 | | | | | 0.31 |
| | | Good | D | E | F | | | | | |
| | | | 0.50 | 0.70 | 0.00 | | | | | 0.40 |

From table 2, it can be observed that student 1 has highest score in excellent category for engineering field, students 2 has highest score in excellent category for medical field and student 3 has highest score in excellent category for hospitality field. This provides some assistance to a student about the career of his capabilities. Here the rules have been framed for three fields only on example basis. Similarly, a detail analysis can be done by including various subjects and vocational interest and defining rules for different fields like commerce, science, literature etc.

Here it is important to note that the fuzzy rules evaluate a student in a holistic manner for a particular field. This is not possible by merely going through the marks in a subject and finding the average of the marks.

### 2.3 Phase 2

After assisting a student the field of his capabilities, in this phase the overall career aspect will be provided by showing him/her the various courses related to the field and institutes within a certain area. Further various jobs related to the courses will be highlighted so that a student will get clear idea about his/her career. For this phase, mathematical modeling can be done as follows:

Let $X = \{x : x \text{ is a career field}\}$, $Y = \{y : y \text{ is an academic course}\}$
$Z = \{z : z \text{ is a job}\}$, $U = \{u : u \text{ is the set of educational institutes}\}$
A relation $R$ from $X$ to $Y$ is defined as
$R = \{(x, y) : x \in X \text{ and } y \in Y \text{ and } y \text{ is an academic course related to the career field } x \}$.
A relation $S$ from $Y$ to $U$ is defined as
$S = \{(y, u) : y \in Y \text{ and } u \in U \text{ and } u \text{ is an educational institute that offers the course } y \}$.
A relation $T$ from $Y$ to $Z$ is defined as
$T = \{(y, z) : y \in Y \text{ and } z \in Z \text{ and } z \text{ is a job related to the course } y \}$.

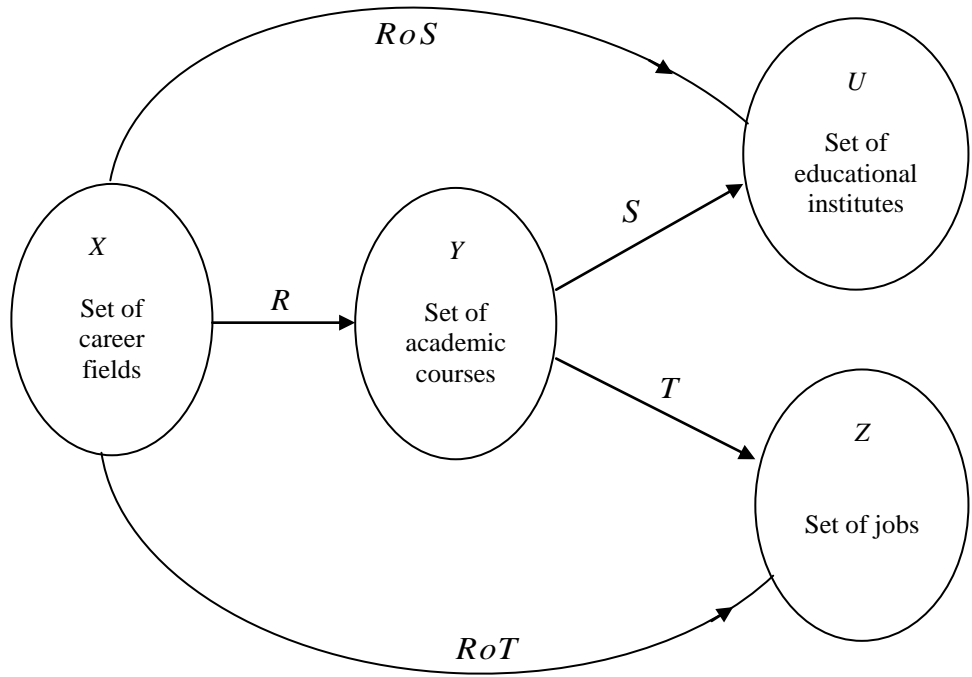

Figure 2. Diagrammatic representation of phase 2 as relations and their compositions

From the above relations, clearly the composition $RoS$ provides the list of institutes related to a career field and the composition $RoT$ provides the list of jobs related to a career field. For a particular career field $x_1 \in X$, the Range ($RoS$) and Range ($RoT$) provide the list of institutes from where an academic course related to a the career field $x_1$ can be done and the jobs that are available after doing the academic course related to career field $x_1$ respectively.

This will provide a complete guideline to a student for his/her career. Figure 2 helps visualize the concept.

## 3. CONCLUSIONS

In the present work a conceptual framework for career counselling has been proposed. The present work proposes a model that analyzes a student based on his marks in various subjects in 10+2 standard and analyzing the vocational interest of a student. Concept of fuzzy inference system provides us the freedom of choosing rules according to our needs. In this paper, rules have been defined for three fields only as an example basis. Similar rules can be created for other fields. Demonstrations of these rules have been done by taking few examples. These examples prove the utility of the system as many times a student is not able to exactly interpret his ability. Combining the vocational interest and subjective interest is quite useful for various filed like hospitality where both capabilities are required. After defining rules for various fields and analyzing them, the system will help a student to know the field of his/her ability.

The second phase of the system will provide a student a clear road map of his/her future, that is, available courses, list of institutes that offer the course and available jobs according to the capabilities of the student. This is an important aspect as a clear road map to future motivates a student to work hard. This will help students to achieve their target without wasting their precious time. Thus the proposed conceptual frame work is quite useful and it will have a real worth when converted to a working system.